\documentclass[12pt,aip]{article}
\begin{document}
\newcommand{\gt}{\mathbf v}
\newcommand{\BB}{\mathbf B}
\newcommand{\jt}{\mathbf j}
\newcommand{\EE}{\mathbf E}
\newcommand{\gto}{{\mathbf v}_{0}}
\newcommand{\BBo}{{\mathbf B}_{0}}
\newcommand{\gte}{{\mathbf v}_{1}}
\newcommand{\BBe}{{\mathbf B}_{1}}
\newcommand{\cte}{{\mathbf j}_{1}}
\newcommand{\AAe}{{\mathbf A}_{1}}
\newcommand{\az}{\mathbf a}
\newcommand{\bz}{\mathbf b}

\title{On existence of resistive magnetohydrodynamic equilibria}

\author{H. Tasso\footnote{het@ipp.mpg.de}, G. N.
Throumoulopoulos\footnote{gthroum@cc.uoi.gr} \\
$^\star$
Max-Planck-Institut f\"{u}r Plasmaphysik \\
Euratom Association \\  85748 Garching bei M\"{u}nchen,
Germany \\ $^\dag$ University of Ioannina,\\ Association Euratom-Hellenic
Republic,\\ Department of Physics, GR 451 10 Ioannina, Greece}
\date{January 24, 2005}
\maketitle
\newpage
\begin{abstract}

A necessary condition for existence of general dissipative magnetohydrodynamic
equilibria is derived. The ingredients of the derivation are Ohm's law and the
existence of magnetic surfaces, only in the sense of KAM theorem. All other
equations describing the system matter exclusively for the evaluation of the
condition in a concrete case.

\end{abstract}
\newpage

Conditions for existence of long time equilibria of hot magnetically confined
plasmas is an important question in fusion research. Such plasmas are weakly
collisional and obey kinetic-Maxwell equations. Conditions for existence of such
equilibria are not known neither within the frame of neoclassical theory nor in
resistive magnetohydrodynamics (MHD). Since neoclassical theory should match
with classical theory in the highly collisional limit, it is of interest to
derive long time equilibrium conditions for the macroscopic equations. The
modest purpose of this note is to derive necessary equilibrium conditions within
the frame of resistive viscous MHD

The basic equations for resistive viscous MHD equilibria
are (see e.g. Ref.\cite{spi})

\begin{eqnarray}
\rho \gt \cdot \nabla \gt
 & = & \jt \times \BB - \nabla p
- \nu \nabla \times \nabla \times \gt + source(momentum) , \\
\EE + \gt \times \BB & = & \mbox{\boldmath $\cal \eta$}\cdot \jt +
\frac{\lambda}{\rho} (\jt \times
\BB - \nabla p_{e}), \\
\nabla \cdot \rho\gt & = & source(mass), \\
\nabla \cdot \BB & = & 0, \\
\nabla \times \EE & = & 0, \\
\nabla \times \BB & = & \jt ,
\end{eqnarray}
where $\rho$ is the mass density, $\nu$ is the viscosity and
$\mbox{\boldmath $\eta$}$ the anisotropic resistivity. p is the total pressure,
$p_{e}$ is the
pressure of the electron fluid, $\gt$ and $\BB$ are the velocity and
the magnetic field, $\EE$ and $\jt$ are the electric field and the
electric current density.

Since only Ohm's law (2) and equations (4)-(5) will be used in the derivation,
the other equations can be modified to accomodate for additional physical
sophistication.
Another important ingredient is the existence of a volume bounded by magnetic
surfaces. Due to Kolmogorov-Arnold-Moser (KAM) theorem (see e.g. Ref.\cite{kam})
this is a mild assumption even for three dimensional equilibria.

Let us take the scalar product of Eq.(2) with the magnetic field to obtain
\begin{equation}
\BB\cdot\nabla\Phi  =\BB \cdot\mbox{\boldmath $\cal \eta$} \cdot \jt -
\frac{\lambda}{\rho}\BB\cdot\nabla p_{e},
\end{equation}
where $\EE$ has been replaced by $\nabla\Phi$ because of Eq.(5). We integrate
now throughout any volume bounded by two magnetic surfaces. By Gauss theorem and
Eq.(4) the gradient term on the RHS of Eq.(7) vanishes if the electron fluid is
barotropic, $p_{e}(\rho)$, or if the electron temperature is constant along the
magnetic field lines. The integral of the first term on the LHS either
vanishes if the
potential $\Phi$ is single valued or is proportional to the loop voltage if the
potential is multi valued.

\begin{itemize}
\item {\bf Tokamak case}: Since the transformer produces a toroidal loop
voltage, the above integration leads to
\begin{equation}
F V_{L} = \int_{KAM}d\tau\BB\cdot\mbox{\boldmath $\cal \eta$}\cdot \jt,
\end{equation}
where $F$ is the toroidal magnetic flux, $V_{L}$ the toroidal loop voltage and
$d\tau$ the volume element.

\item {\bf Stellarator case}: Since no electric field is induced, the potential
is
single valued, so the above integration leads to
\begin{equation}
\int_{KAM}d\tau\BB\cdot\mbox{\boldmath $\cal \eta$}\cdot \jt = 0.
\end{equation}
\end{itemize}

Conditions (8) and (9) are only necessary since they do not need the full system
(1)-(6) for their derivation. They give, however, a qualitative insight of how
the current density has to behave in order to fulfil the stationarity condition.
The evaluation of the nonlocal conditions (8) and (9) needs the knowledge of the
KAM surfaces in the three dimensional MHD equilibrium as well as the resistivity
tensor, the magnetic field and the current density. For many axisymmetric
equilibria the magnetic surfaces are known as well as the magnetic field and the
current density. In the latter case one can find local conditions
for existence of resistive MHD incompressible equilibria and even solve the full
system (1)-(6) as demonstrated in Ref.\cite{tt}. The
nonlocal conditions (8) and (9) apply, however, to any toroidal MHD equilibrium
without restricting the number of dimensions by symmetry.

\newpage

\begin{center}
{\large\bf Acknowledgements}
\end{center}

Part of this work was conducted during a visit of one of the authors (G.N.T.) to
the Max-Planck-Institut f\"{u}r Plasmaphysik, Garching. The hospitality of that
Institute is greatly appreciated.
The present work was performed under the Contract of Association ERB 5005 CT 99
0100 between the European Atomic Energy Community and the Hellenic Republic.

\end{document}